\documentclass[%
 reprint,
 onecolumn,
 amsmath,amssymb,
 aps,
]{revtex4-1}

\usepackage{graphicx}
\usepackage{dcolumn}
\usepackage{bm}
\usepackage{subfigure}
\usepackage{xcolor}

\def\bc{\begin{center}}
\def\ec{\end{center}}
\def\be{\begin{eqnarray}}
\def\ee{\end{eqnarray}}

\begin{document}

\preprint{APS/123-QED}

\begin{center}
\large{\textbf{CosmoMC Installation and Running Guidelines}}
\end{center}

\author{Ming-Hua Li}
  \email{395149052@qq.com}
\affiliation{%
 School of Physics and Engineering, Sun Yat-Sen University, 510275 Guangzhou, China
}%

\author{Ping Wang}
  \email{pwang@ihep.ac.cn}
\affiliation{%
Division for Particle Astrophysics, Institute of High Energy Physics, Chinese Academy of Sciences, 100049 Beijing, China
}%

\author{Zhe Chang}
  \email{changz@mail.ihep.ac.cn}
\author{and Dong Zhao}
\email{zhaod@ihep.ac.cn}
\affiliation{%
Institute of High Energy Physics, Chinese Academy of Sciences, 100049 Beijing, China}
\altaffiliation[Also at ]{Theoretical Physics Center for Science Facilities, Chinese Academy of Sciences, 100049 Beijing, China
}%

\begin{abstract}
CosmoMC\footnote{\url{http://cosmologist.info/cosmomc/}. In this guide, we use the CosmoMC released on July 23, 2014.} is a Fortran 95 Markov-Chain Monte-Carlo (MCMC) engine to explore the cosmological parameter space, plus a Python suite for plotting and presenting results. This document describes the installation of the CosmoMC on a Linux system\footnote{This guidebook is based on the operating system (OS) Ubuntu 14.04.1 LTS 64-bit version. For other versions of Linux, the procedures are almost the same.}. It is written for those who want to use it in their scientific research but without much training on Linux and the program. Besides a step-by-step installation guide, we also give a brief introduction of how to run the program on both a desktop and a cluster. We share our way to generate the plots that are commonly used in the cosmological references. For more information, one can refer to the CosmoCoffee\footnote{CosmoCoffee: \url{http://cosmocoffee.info/viewforum.php?f=11}} forum or contact the authors of this document. Questions and comments would be much appreciated. \\
\end{abstract}

\maketitle
\newpage
$~$\\
$~$\\
$~$\\
$~$\\
$~$\\
$~$\\
$~$\\
$~$\\
$~$\\
$~$\\
$~$\\
$~$\\
$~$\\
$~~~~~~~~~~~~~~~~~~~~~~~~~~~~~~~~~~~~~~~~~~~~~~${\Large Learn, Discuss, and Contribute}

\newpage
For a successful installation and running of CosmoMC, the following prerequisites are required:
\begin{itemize}
   \item Intel@Fortran Compiler (version 13$+$) $~~~$(\url{https://software.intel.com/en-us/fortran-compilers})
   \item Open MPI $~~~~~~~~~~~~~~~~~~~~~~$(\url{http://www.open-mpi.org})
   \item CFITSIO $~~~~~~~~~~~~~~~~~~~~~~~$(\url{http://heasarc.gsfc.nasa.gov/docs/software/fitsio/fitsio.html})
   \item HEALPix $~~~~~~~~~~~~~~~~~~~~~~~$(\url{http://healpix.sourceforge.net})
   \item WMAP or Planck likelihood data $~~~$(\url{http://lambda.gsfc.nasa.gov/product/map/dr5/likelihood_get.cfm})
\end{itemize}

The rest of the document is organized as follows. In Section I, we guide you to install the prerequisites and likelihood data that are necessary for a successful compilation of CosmoMC. In Section II, we give a brief introduction of how to run the program and the way to generate the plots that one usually sees in the references of cosmology. FAQs about running the program are also presented. \\

\section{Installation and Compilation}
{\bf Intel@Fortran Compiler} (version 13 or a more recent release) and {\bf Open MPI} are both necessary for a successful compilation of the CosmoMC package. Intel@Fortran Compiler should be installed prior to the Open MPI. Before installing these prerequisites, you need to do some presettings of your Linux system.\\
\\
1. Set the password for root user. \\
$~~~~~~$\texttt{\$ sudo passwd} \\
$~~~~~~$\texttt{\$ [sudo] password for Your\_User\_Name:} (type in the password you input during the installation of the system) \\
$~~~~~~$\texttt{\$ Enter new UNIX password:} (input the new password for the root user) \\
$~~~~~~$\texttt{\$ Retype new UNIX password:} (retype the new password)\\
$~~~~~~$\texttt{\$ password:} password updated successfully\\
2. Run the updates before installing any new packages. \\
$~~~~~~$\texttt{\$ sudo apt-get update} \\
3. Have the g++ package installed.\\
$~~~~~~$\texttt{\$ sudo apt-get install g++}  \\
4. Install the possibly essential packages you will need in your further building of other software packages.\\
$~~~~~~$\texttt{\$ sudo apt-get install build-essential}  \\

\subsection{Installing Intel@Fortran Compiler}
CosmoMC needs Intel@Fortran Compiler (version 13 or a more recent release) to compile. Having your OS installed and preset, now we can set out to install the Intel@Fortran Compiler. A better choose is to install the {\bf Intel@Fortran Composer XE 2013} (Intel@Fortran Compiler is a part of it.) See the following guidelines for installation \citep{ifort}. My home path is {\bf /home/limh}. All the source files are placed in the directory {\bf /home/limh/Downloads}. All the softwares are installed into the directory {\bf /home/limh/Programs}. My working directory is {\bf /home/limh/workspace}. You can use your own settings during the installation.\\
\\
\noindent
1. Go to \url{https://software.intel.com/en-us/non-commercial-software-development} to download a non-commercial version of Intel@Fortran Composer XE 2013 for Linux.\\
2. After filling a form, you will have a non-commercial serial number sent to your registered email. Use it to register and download the source file {\bf l\_fcompxe\_2013\_sp1.3.174.tgz}. \\
3. Go to the Terminal and untar the file:\\
$~~~~~~$\texttt{\$ tar -xzvf l\_fcompxe\_2013\_sp1.3.174.tgz} \\
4. Run the {\bf install.sh} file to start the installation. Before doing this, you may have the serial number ready.\\
$~~~~~~$\texttt{\$ cd l\_fcompxe\_2013\_sp1.3.174}  \\
$~~~~~~$\texttt{\$ ./install.sh}\\
Here I provide a noncommercial serial number used by myself, i.e. NTJL-484Z4TT5. It will be expired by July 19, 2015. You can then apply for another one. In the step 5 of 7, please remember to change the install directory as {\bf (your own installing path)/intel/composer\_xe\_2013\_sp1.3.174}.\\
5. Set the environment variable. Enter the home directory (for me, this is {\bf /home/limh}) and run:  \\
$~~~~~~$\texttt{\$ vim .bashrc}  \\
$~~~~$Add the following command line at the end of the {\bf .bashrc} file \citep{sourceifort}: \\
$~~~~~~~~$\texttt{source /home/limh/Programs/intel/composer\_xe\_2013\_sp1.3.174/bin/ifortvars.sh intel64}\\
$~~~~$For a 32-bit Linux system, use the following line instead:\\
$~~~~~~~~$\texttt{source /home/limh/Programs/intel/composer\_xe\_2013\_sp1.3.174/bin/ifortvars.sh ia32}\\
$~~~~$(Press Esc and type in `\texttt{:wq}' to save and exit.)\\
6. Restart the terminal and run: \\
$~~~~~~$\texttt{\$ ifort --version  \\
$~~~~~~$ifort (IFORT) 14.0.3 20140422 \\
$~~~~~~$Copyright (C) 1985-2014 Intel Corporation. All right reserved.}\\
$~~~~$Seeing the above outputs means that Intel@Fortran Composer XE 2013 has already been successfully installed on your Linux. To make sure the command {\bf ifort} works, please remember to run the command {\bf sourtce .bashrc} every time you restart the terminal.\\

\subsection{Installing Open MPI}
After having the Intel@Fortran Compiler installed, we can set out to install the Open MPI. The installation may take almost 40 minutes.\\
1. Go to \url{http://www.open-mpi.org/software/ompi/v1.8/} to download the newest stable version. This installation is based on the Version 1.8.1.\\
2. Untar the source file using\\
$~~~~~~$\texttt{\$ tar -zvxf openmpi-1.8.1.tar.gz} \\
3. Configure and make. Do remember to set the {\bf F77}, {\bf FC} and {\bf F90} environment variables before configuring (for one to use Intel@Fortran Compiler instead of Gfortran or other compilers. Here, I use a 64-bit version of Intel@Fortran Compiler), e.g. \\
$~~~~~~$\texttt{\$ cd openmpi-1.8.1} \\
$~~~~~~$\texttt{\$ ./configure --prefix=/home/limh/Programs/openmpi$~~$F77=/home/limh/Programs/intel/}\\
$~~~~~~~~~~~~~$\texttt{composer\_xe\_2013\_sp1.3.174/bin/intel64/ifort$~~$FC=/home/limh/Programs/intel/composer\_xe\_2013\_sp1.3.174}\\
$~~~~~~~~~~~~$\texttt{/bin/intel64/ifort$~~$F90=/home/limh/Programs/intel/composer\_xe\_2013\_sp1.3.174/bin/intel64/ifort} \\
$~~~~~~$\texttt{\$ make} \\
$~~~~~~$\texttt{\$ make install} \\
4. Go to the home directory (for me this is {\bf /home/limh}) and add the following lines at the end of the {\bf .bashrc} file: \\
$~~~~~~$\texttt{\$ vi .bashrc} \\
$~~~~$Add the following line at the end of the file. Quit and save. \\
$~~~~~~~~$ \texttt{export PATH=/home/limh/Programs/openmpi/bin:\$\{PATH\}} \\
$~~~~$Then run {\bf source .bashrc} to implement the bash settings.\\
$~~~~~~$\texttt{\$ source .bashrc} \\
5. Enter the {\bf openmpi-1.8.1/examples/} directory of Open MPI to run the test.\\
$~~~~~~$\texttt{\$ cd /scratchfs/hnlin/openmpi-1.8.1/examples} \\
$~~~~$Save and quit. Then run {\bf make} to compile. \\
$~~~~~~$\texttt{\$ make} \\
$~~~~~~$\texttt{\$ mpirun  -np 4 ./hello\_c} \\
$~~~~$Seeing the following outputs means that Open MPI has already been successfully installed on your Linux. \\
$~~~~~~$\texttt{Hello, world, I am 0 of 4, (Open MPI v1.8.1, package: Open MPI limh@limh-pc Distribution, ident: 1.8.1, Apr 22, 2014, 89)}\\
$~~~~~~$\texttt{Hello, world, I am 1 of 4, (Open MPI v1.8.1, package: Open MPI limh@limh-pc Distribution, ident: 1.8.1, Apr 22, 2014, 89)}\\
$~~~~~~$\texttt{Hello, world, I am 3 of 4, (Open MPI v1.8.1, package: Open MPI limh@limh-pc Distribution, ident: 1.8.1, Apr 22, 2014, 89)}\\
$~~~~~~$\texttt{Hello, world, I am 2 of 4, (Open MPI v1.8.1, package: Open MPI limh@limh-pc Distribution, ident: 1.8.1, Apr 22, 2014, 89)}\\

\subsection{Installing CFITSIO}
CFITSIO is also another prerequisite for one to run CosmoMC. Run the following command lines to install it.\\
1. Untar, configure, and make (others also use `{\bf make shared}' instead of `{\bf make}'). \\
$~~~~~~$\texttt{\$ tar -xzvf cfitsio3370.tar.gz}  \\
$~~~~~~$\texttt{\$ cd cfitsio} \\
$~~~~~~$\texttt{\$ ./configure - -prefix=/home/limh/Programs/cfitsio} \\
$~~~~~~$\texttt{\$ make} \\
$~~~~~~$\texttt{\$ make install} \\
2. Go to the home directory (for me this is {\bf /home/limh}) and add the following lines at the end of the {\bf .bashrc} file: \\
$~~~~~~$\texttt{\$ vi .bashrc} \\
$~~~~$Add the following line at the end of the file. Quit and save. \\
$~~~~~~~~$ \texttt{export LD\_LIBRARY\_PATH=/home/limh/Programs/cfitsio/lib:\$\{LD\_LIBRARY\_PATH\}} \\

\subsection{Installing HEALPix}
The installation of HEALPix is a little bit bothersome. One can build the facilities with a number of compilers, i.e. C, C++, Fortran, Python, etc. In this guidebook, we use the Fortran compiler approach.\\
1. Untar the source file and run {\bf ./configure}. For the untar directory is just the installation directory, it is suggested that you copy the source file into the directory {\bf /home/limh/Programs/} before you untar it. \\
$~~~~~~$\texttt{\$ cp /home/limh/Downloads/Healpix\_3.11\_2013Apr24.tar.gz /home/limh/Programs/} \\
$~~~~~~$\texttt{\$ cd /home/limh/Programs} \\
$~~~~~~$\texttt{\$ tar -zvxf Healpix\_3.11\_2013Apr24.tar.gz} \\
$~~~~~~$\texttt{\$ ./configure} \\
$~~~~$After doing this, you will see a menu with several options. Choose the Fortran compiling options and then follow the self-explanatory guidelines it gives. Do not edit the default compiling options and settings for the compilers if you don't know what it means. You will be asked to give the name of your Fortran compiler (just type in {\bf ifort}) and the installation path of cfitsio (i.e. {\bf /home/limh/Programs/cfitsio}). When it goes back to the starting menu, type in `{\bf 0}' and then press Enter to exit. \\
2. Run `make' to build the facilities. \\
$~~~~~~$\texttt{\$ make} \\
$~~~~~~$\texttt{\$ make test} \\
$~~~~$Seeing a successful message means you have successfully installed HEALPix.\\
$~~~~~~~~$\texttt{...}\\
$~~~~~~~~$\texttt{process\_mask> normal completion}\\
$~~~~~~~~$\texttt{Healpix F90 tests done} \\
$~~~~~~~~$\texttt{success rate: 10/10} \\

\subsection{Building WMAP Likelihood Data}
\noindent 
1. Go to \url{http://lambda.gsfc.nasa.gov/product/map/dr5/likelihood_get.cfm} to download the source file {\bf wmap\_likelihood\_full\_v5.tar.gz}.\\
2. Copy the file to the working directory (for me, this is \texttt{/home/limh/workspace/}). Untar the source file.\\
$~~~~~~$\texttt{\$ cp /home/limh/Downloads/wmap\_likelihood\_full\_v5.tar.gz /home/limh/workspace} \\
$~~~~~~$\texttt{\$ tar -zvxf wmap\_likelihood\_full\_v5.tar.gz} \\
3. Edit the Makefile. Add the installation path of cfitsio and MKL library of the intel Fortran compiler (the modifications are shown in {\bf boldface}). \\
$~~~~~~$\texttt{\$ cd wmap\_likelihood\_full\_v5} \\
$~~~~~~$\texttt{\$ vi Makefile} \\
$~~~~~~~~$\texttt{...}\\
$~~~~~~~~$\texttt{CFITSIO={\bf /home/limh/Programs/cfitsio}} \\
$~~~~~~~~$\texttt{...}\\
$~~~~~~~~$\texttt{\# Linux/Intel compiler and MKL libraries} \\
$~~~~~~~~$\texttt{{\bf MKLPATH = /home/limh/Programs/intel/composer\_xe\_2013\_sp1.3.174/mkl/lib/intel64}}\\
$~~~~~~~~$\texttt{F90 = ifort} \\
$~~~~~~~~$\texttt{FFLAGS = -02 -fpic \$(WMAPFLAGS)}\\
$~~~~~~~~$\texttt{INCS = -I. -I\$(CFITSIO)/include} \\
$~~~~~~~~$\texttt{LIBS = -L. -L{\bf \$(MKLPATH)} -lmkl\_intel\_lp64 -lmkl\_intel\_thread -lmkl\_core -liomp5 -lmkl\_mc3} \\
$~~~~~~~~~~~~~~~~~~~~$\texttt{-lmkl\_def {\bf -lmkl\_lapack95\_lp64} -L\$(CFITSIO){\bf /lib} -lcfitsio} \\
$~~~~~~~~$\texttt{...}\\
4. Open the {\bf WMAP\_9yr\_options.f90} file. Modify the first line to give the path of the data directory of wmap\_likelihood\_full\_v5 (the modifications are shown in {\bf boldface}).\\
$~~~~~~$\texttt{\$ vim WMAP\_9yr\_options.f90} \\
$~~~~~~~~$\texttt{...}\\
$~~~~~~~~$\texttt{Character(Len=128) :: WMAP\_data\_dir = `{\bf /home/limh/workspace/wmap\_likelihood\_v5/data/}' }\\
$~~~~~~~~$\texttt{...}\\
5. Run {\bf make all} to build wmap\_likelihood\_full\_v5.\\
$~~~~~~$\texttt{\$ make all} \\

\subsection{Building Planck Data}
If you want to use the Planck data in your study, then you have to build it as the WMAP likelihood data. The procedure is somewhat different \citep{planckreadme}.\\
1. Go to \url{http://cosmologist.info/cosmomc/readme_planck.html}. Use the hyperlink it provides to download the source file {\bf  plc-1.0.tar.gz}. \\
2. Copy the file to the working directory (for me, this is {\bf /home/limh/workspace/}). Untar the source file.\\
$~~~~~~$\texttt{\$ cp /home/limh/Downloads/plc-1.0.tar.gz /home/limh/workspace} \\
$~~~~~~$\texttt{\$ tar -xzvf plc-1.0.tar.gz}  \\
3. Go to the directory and edit the {\bf Makefile} to give the library path of cfitsio and that of the Intel@Fortran compiler (the modifications are shown in {\bf boldface}).  \\
$~~~~~~$\texttt{\$ cd plc-1.0} \\
$~~~~~~$\texttt{\$ vi Makefile} \\
$~~~~~~~~$\texttt{...}\\
$~~~~~~~~$CFITSIOPATH = {\bf /home/limh/Programs/cfitsio} \\
$~~~~~~~~$\texttt{...}\\
$~~~~~~~~$\texttt{\# on a linux machine, ifort 11.1} \\
$~~~~~~~$\texttt{ IFORTLIBPATH = {\bf /home/limh/Programs/intel/composer\_xe\_2013\_sp1.3.174/compiler/lib}} \\
$~~~~~~~$\texttt{ IFORTRUNTIME = -L\$(IFORTLIBPATH) -lintlc -limf -lsvml -liomp5 -lifport -lifcoremt -lpthread }\\
$~~~~~~~~$\texttt{...}\\
$~~~~$Here you can either edit the {\bf MKLROOT} and {\bf LAPACKLIBPATHMKL} to give the installation path of the MKL library (you can run {\bf echo \$MKLROOT} to find your own installation path), e.g.\\
$~~~~~~~~$\texttt{...}\\
$~~~~~~~~$MKLROOT = {\bf /home/limh/Programs/intel/composer\_xe\_2013\_sp1.3.174/mkl} \\
$~~~~~~~~$LAPACKLIBPATHMKL = {\bf -L\${MKLROOT}/lib/intel64} \\
$~~~~~~~~$\texttt{...}\\
or just  leave it empty to be set during {\bf configure} using the options {\bf --lapack\_mkl=\${MKLROOT}}. \\
4. Configure and make. See the instructions on the webpage \url{http://cosmologist.info/cosmomc/readme_planck.html}. If you have already set the installation path of the MKL library (i.e. editted the {\bf MKLROOT} and {\bf LAPACKLIBPATHMKL} variables in the Makefile), then you should omit the option {\bf - -lapack\_mkl=\${MKLROOT}} in the following commands. Otherwise, you should just run the following commands: \\
$~~~~~~$\texttt{\$ cd plc-1.0} \\
$~~~~~~$\texttt{\$ ./waf configure --lapack\_mkl=\${MKLROOT} --lapack\_mkl\_version=10.3 --install\_all\_deps} \\
$~~~~~~$\texttt{\$ ./waf install} \\
5. Add the following lines into the {\bf .bashrc} file in the home directly (for me, this is {\bf /home/limh}). \\
$~~~~~~$\texttt{\$ vi /home/limh/.bashrc} \\
$~~~~~~~~$\texttt{...}\\
$~~~~~~~~$\texttt{ export PLANCKLIKE=cliklike }\\
$~~~~~~~~$\texttt{ export CLIKPATH=/home/limh/workspace/plc-1.0 }\\
$~~~~~~~~$\texttt{ export LD\_LIBRARY\_PATH=\$LD\_LIBRARY\_PATH:\$CLIKPATH/lib }\\
$~~~~$Save and quit. Then run {\bf  source /home/limh/.bashrc} to implement the bash settings.\\
6. Go the cosmomc root directory (for me, this is {\bf /home/limh/workspace/cosmomc}) and make a static symbolic link to the Planck data. \\
$~~~~~~$\texttt{\$ cd /home/limh/workspace/cosmomc} \\
$~~~~~~$\texttt{\$ ln -s /home/limh/workspace/plc-1.0 ./data/clik} \\

\subsection{Compiling CosmoMC}
Having all the prerequisites installed, now we can set out to compile and run CosmoMC. Note that the compile of CosmoMC is carried out in the {\bf cosmomc/source} directory. \\
1. Go to \url{http://cosmologist.info/cosmomc/readme.html}. Use the hyperlink it provides to download the source file {\bf cosmomc.tar.gz}.\\ 
2. Untar the source file `cosmomc.tar.gz'. \\
$~~~~~~$\texttt{\$ cp /home/limh/Downloads/cosmomc.tar.gz /home/limh/workspace} \\
$~~~~~~$\texttt{\$ tar -xzvf cosmomc.tar.gz}  \\
3. Edit the {\bf Makefile} in the directory {\bf cosmomc/source/} to give the installation path of cfitsio (the modifications are shown in {\bf boldface}). \\
$~~~~~~$\texttt{\$ cd cosmomc/source} \\
$~~~~~~$\texttt{\$ vi Makefile} \\
$~~~~~~~~$\texttt{...}\\
$~~~~~~~~$\texttt{WMAP = {\bf /home/limh/workspace/wmap\_likelihood\_v5}} \\
$~~~~~~~~$\texttt{...}\\
$~~~~~~~~$\texttt{cfitsio = {\bf /home/limh/Programs/cfitsio}} \\
$~~~~$Here remember not to edit the {\bf PLANCKLIKE} environment variable content even if one wants to use the Planck data in CosmoMC. Because we have already set this in the {\bf .bashrc} file when we build the Planck data. We don't have to do it again in the Makefile of CosmoMC. \\
$~~~~~~~~$\texttt{...}\\
$~~~~~~~~$\texttt{ifeq (\$(COSMOHOST), darwin) }\\
$~~~~~~~~$\texttt{WMAP = {\bf /home/limh/workspace/wmap\_likelihood\_v5} }\\
$~~~~~~~~$\texttt{cfitsio = {\bf /home/limh/Programs/cfitsio} }\\
$~~~~~~~~$\texttt{endif} \\
4. Edit the {\bf Makefile} in the directory {\bf cosmomc/camb/} to give the installation path of cfitsio and Healpix (the modifications are shown in {\bf boldface}). \\
$~~~~~~$\texttt{\$ cd /home/limh/workspace/cosmomc/camb} \\
$~~~~~~$\texttt{\$ vi Makefile} \\
$~~~~~~~~$\texttt{...}\\
$~~~~~~~~$\texttt{FITSDIR = {\bf /home/limh/Programs/cfitsio}} \\
$~~~~~~~~$\texttt{FITSLIB = {\bf /home/limh/Programs/cfitsio/lib}} \\
$~~~~~~~~$\texttt{HEALPIXDIR = {\bf /home/limh/workspace/Healpix\_3.11}} \\
5. Run `make' and `make getdist'. \\
$~~~~~~$\texttt{\$ make} \\
$~~~~~~$\texttt{\$ make getdist} \\
$~~~~$After running {\bf make getdist}, you will find that  a file named {\bf getdist} generated in the cosmomc root directory. Then you can go to the cosmomc root directory to run CosmoMC. \\
$~~~~$If you get error message like `{\bf libimf.so: warning: feupdateenv is not implemented and will always fail}', just add `{\bf -limf}' (or `{\bf -limf -lm}') to the {\bf FFLAGS} (options of the compiler) in the {\bf Makefile} in the directory {\bf /home/limh/workspace/cosmomc/source}. It tells the compiler to link not only the math libraries of the Intel compilers but also those of the system \citep{libimf}. The Makefile would look like\\
$~~~~~~~~$\texttt{...}\\
$~~~~~~~~$\texttt{F90C = ifort} \\
$~~~~~~~~$\texttt{FFLAGS = -openmp -fast -w -fpp2 {\bf -limf}} \\
$~~~~~~~~$\texttt{...}\\

\section{Running CosmoMC and Generating Plots}
\subsection{On a Desktop}
Every time you modify the code, it is recommended that you first `{\bf make}' the CosmoMC on your own desktop to make sure it there is no errors or warnings during the compile. You can use the {\bf test.ini} or {\bf params\_generic.ini} provided in the root directory of cosmomc to do this. A detailed description of the files and folders in the root directory of CosmoMC is given in the {\bf readme.html} file in its root directory, which is also available on the website \url{http://cosmologist.info/cosmomc/readme.html}. \\
1. Run and test CosmoMC. You can use {\bf ./cosmomc test.ini} or {\bf mpirun -np 2 ./cosmomc params\_generic.ini} to do this. \\
$~~~~~~$\texttt{\$ cd /home/limh/workspace/cosmomc} \\
$~~~~~~$\texttt{\$ mpirun -np 2 ./cosmomc params\_generic.ini} \\
$~~~~$This command will generate data file in the {\bf /cosmomc/chains} directory. For a successful convergent run, you can see the time it takes from the feedback message in the Terminal, e.g. \\
$~~~~~~~~$\texttt{ ...}\\
$~~~~~~~~$\texttt{Chain $~~~~~$ 2 $~~$ MPI $~$ Communicating} \\
$~~~~~~~~$\texttt{Chain $~~~~~$ 1 $~~$ MPI $~$ Communicating} \\
$~~~~~~~~$\texttt{Current convergence R-1 =  $~~$ 5.4434482E-03 $~~~$chain steps = $~~$ 437} \\
$~~~~~~~~$\texttt{Requested convergence R achieved} \\
$~~~~~~~$\texttt{Total time: $~~$ 0 $~~$ ( $~~$ 0.00011 hours $~~$)}\\
$~~~~$Then one can run {\bf mpirun -np 2 ./getdist distgeneric.ini} to generate the plot data. Before doing so, one has to manually create a folder named {\bf plot\_data} in the root directory of CosmoMC (for me, this is {\bf /home/limh/workspace\\
/cosmomc}). \\
2. Create the plot\_data folder and generate the plot data. \\
$~~~~~~$\texttt{\$ cd /home/limh/workspace/cosmomc} \\
$~~~~~~$\texttt{\$ mkdir plot\_data} \\
$~~~~~~$\texttt{\$ mpirun -np 2 ./getdist distgeneric.ini} \\
3. Use Python to generate the plots. Before doing this, one has to copy all the files in the {\bf cosmomc/python/} directory (most essential the {\bf GetDistPlot.py}) to the root directory of CosmoMC (for me, it is {\bf /home/limh/workspace\\
/cosmomc}). \\
$~~~~~~$\texttt{\$ cp /home/limh/workspace/cosmomc/python/*.* /home/limh/workspace/cosmomc} \\
4. Then you can go to the Ubuntu Software Center to download and install the {\bf IDLE(using Python-2.7)} (an integrated development environment for python using Python-2.7). \\
(1) After successfully installing IDLE(using Python-2.7), click the icon to start it. It will start a new window titled `Python 2.7.6 Shell'.\\
(2) Click `{\bf File} $\rightarrow$ {\bf Open}', go to the directory {\bf /home/limh/workspace/cosmomc} and choose {\bf test.py} (or {\bf test\_tri.py}) to open. Then another window will pop out, displaying the content of the `test.py' file (or `test\_tri.py' file). \\
(3) Then press {\bf F5} on the keyboard to run the script. If everything works fine, there would be a pdf file named {\bf test.pdf} (or {\bf test\_tri.pdf}) generated in the directory {\bf /home/limh/workspace/cosmomc}. That is the parameter plot one expects from the CosmoMC.

\subsection{On a Cluster}
A full implementation of CosmoMC needs a cluster environment. To have it run on a cluster, one should first successfully {\bf make} it (see Section II, A. Compiling CosmoMC) on his own desktop. Then one has to write a job-submit script in the root directory of CosmoMC and have it run in the Terminal of the cluster. A {\bf job-submit script} is just a text file written in the bash language telling the cluster how to run the program. It specifies the root directory of CosmoMC, the nodes and cores on which you want to run CosmoMC, and the name and path of the output files in which you want the feedbacks and error messages (if there are any) to be saved, etc. Samples of a {\bf PBS(Protable Batch System)} job-submit script are easy to obtain on the Internet. 

Here we give our job-submit script sample below. It is written as neatly as possible for the beginners. One can copy the file and edit it as appropriate for your own machine. It is named {\bf limhsub.sh} and placed in the directory {\bf /ihepbatch/mbhd01/user/liyangrong/liminghua/limh/cosmomc/} on the cluster.\\
$~$\\
\texttt{\#\#\#\#\#\#\#\#\#\#\#\#\#\#\#\#\#\#\#\#\#\#\#\#\#\#\#\#\#\#\#\#\#\#\#\#\#\#\#\#\#\#\#\#\#\#\#\#\#\#\#\#\#\#\#\#\#\#\#\#\#\#\#\#\#\#\#\#\#\#\#\#\#\#\#\#\#\#\#\#\#\#\#\#}\\
\texttt{\#!/bin/bash} \\
\texttt{\#PBS -N limhsub.sh}  \\
\texttt{\#PBS -o /ihepbatch/mbhd01/user/liyangrong/liminghua/limh/limhsub.out} \\
\texttt{\#PBS -e /ihepbatch/mbhd01/user/liyangrong/liminghua/limh/limhsub.err} \\
\texttt{\#PBS -l nodes=2:ppn=8}  \\
$~$\\
\texttt{DIR\_HERE=/ihepbatch/mbhd01/user/liyangrong/liminghua/limh/cosmomc} \\
\texttt{DIR\_MPI=/ihepbatch/mbhd01/user/liyanrong/liminghua/Programs/openmpi/bin} \\
$~$\\
\texttt{cat \$PBS\_NODEFILE $>$ \$DIR\_HERE/hostfile} \\
\texttt{NCPU=`wc -l $<$ \$PBS\_NODEFILE`} \\
$~$\\
\texttt{source /ihepbatch/mbhd01/user/liyanrong/liminghua/.bashrc} \\
\texttt{cd \$PBS\_O\_WORKDIR} \\
\texttt{\$DIR\_MPI/mpirun -hostfile \$\{PBS\_NODEFILE\} -np \$NCPU \$DIR\_HERE/cosmomc \$DIR\_HERE/params\_generic.ini} \\
$~$\\
$~$\\
\texttt{\#\# The following lines are annotations. \\
\#\# The first line `\#!/bin/bash' is not an annotation. It tells the cluster that this script is written in the bash language. \\
\#\# The second to fifth lines that starts with `\#PBS' are not annotations. They are the command lines of the PBS system.\\
\#\# The fifth line means using 2 nodes to run the CosmoMC. Each node uses 8 cores.\\
\#\# The seventh and eighth lines respectively gives the path of the working directory of CosmoMC and Open MPI. \\
\#\# The tenth line uses the bash command `cat' to show the content of the file `\$PBS\_NODEFILE' and write into a new file named `hostfile' in the root directory of CosmoMC. \\
\#\# The eleventh line counts the number of available nodes and save it in the variable \$NCPU.\\
\#\# The thirteenth line `source ...' is necessary for the cluster to recognize the Intel@Fortran compiler, Open MPI, and the Planck dataset etc.\\
\#\# The fourteenth line is to go to the working directory \$DIR\_HERE. Thus the cluster can identify the directory paths in the codes that are given with respect to the root directory of CosmoMC.\\
\#\# The last line tells the cluster to run the `cosmomc' file with its options-file `params\_generic.ini'. `-np' and `-hostfile' is the options of Open MPI. Appropriate options should be used for specific cluster environments. For example, an Infiniband- or Ethernet-communicating cluster demands different MPI options. See `http://www.open-mpi.org/faq/?category=openfabrics\#ib-btl' for more information. \\
\#\# If you have any troubles in doing this, you can: \\
\#\# (1) Run `mpirun --help' in the Terminal for a brief introduction of the options; \\
\#\# (2) Go to the FAQ of Open MPI mailing lists `http://www.open-mpi.org/faq/?category=openfabrics\#ib-btl' for more information (we find this very helpful); \\
\#\# (3) Contact the system administrator to get more support.\\
}
\texttt{\#\#\#\#\#\#\#\#\#\#\#\#\#\#\#\#\#\#\#\#\#\#\#\#\#\#\#\#\#\#\#\#\#\#\#\#\#\#\#\#\#\#\#\#\#\#\#\#\#\#\#\#\#\#\#\#\#\#\#\#\#\#\#\#\#\#\#\#\#\#\#\#\#\#\#\#\#\#\#\#\#\#\#\#}\\

To submit a job-submit script, one can use the command {\bf qsub}, following the name of the script file. For more information about job submission, just `google' the {\bf Torque} or {\bf PBS} system, or consult with your system administrator for more support.

\subsection{FAQs about Running}
1. (On a cluster) If you have limited registered memory error messages in your {\bf .err} file like:\\
$~~~~~~~~$\texttt{libibverbs: Warning: RLIMIT\_MEMLOCK is 32768 bytes.}\\
$~~~~~~~~$\texttt{This will severely limit memory registrations.}\\
$~~~~~~~~$\texttt{...}\\
and the {\bf .out} file like\\
$~~~~~~~~$\texttt{The OpenFabrics (openib) BTL failed to initialize while trying to
allocate some locked memory.  This typically can indicate that the
memlock limits are set too low.  For most HPC installations, the
memlock limits should be set to "unlimited".  The failure occured
here:\\
$~~$\\
$~~~~~~~~$Local host:    mbh023\\
$~~~~~~~~$OMPI source:   btl\_openib.c:872\\
$~~~~~~~~$Function:      ompi\_free\_list\_init\_ex\_new()\\
$~~~~~~~~$Device:        mlx4\_0\\
$~~~~~~~~$Memlock limit: 32768\\
$~~~~~~~~$...}\\
$~~~~$you may need to consult with your system administrator to get this problem fixed \citep{memorylock}.\\
$~~$\\
2. (On a desktop or cluster) If you get error message like:\\
$~~~~~~~~$\texttt{libimf.so: warning: feupdateenv is not implemented and will always fail}\\
$~~~~~~~~$\texttt{...}\\
$~~~~$just add `{\bf -limf}' (or `{\bf -limf -lm}') to the {\bf FFLAGS} (options of the compiler) in the {\bf Makefile} in the {\bf source} folder in the root directory of CosmoMC (for me, this is {\bf /home/limh/workspace/cosmomc/source}). It tells the compiler to link not only the math libraries of the Intel compilers but also those of the system \citep{libimf}. The Makefile would look like\\
$~~~~~~~~$\texttt{...}\\
$~~~~~~~~$\texttt{F90C = ifort} \\
$~~~~~~~~$\texttt{FFLAGS = -openmp -fast -w -fpp2 {\bf -limf}} \\
$~~~~~~~~$\texttt{...}\\

\section*{Acknowledgments}
We would like to thank Antony Lewis from the University of Sussex for making the CosmoMC code publicly available. We are grateful to Prof. Jian-Min Wang and Yan-Rong Li from the Division for Particle Astrophysics of the Institute of High Energy Physics for allowing us to run and test the code on their machine.
We thank Prof. Xin Li from the Chongqing University (Chongqing 400030, China) and Hai-Nan Lin and Yu Sang from the theory division of the Institute of High Energy Physics for useful discussions about the code. We would also like to thank Prof. Zhi-Bing Li from the School of Physics and Engineering of the Sun Yat-Sen University (Guangzhou 510275, China) for his support of our work.


\end{document}